\documentclass[review]{elsarticle}
\usepackage{amsmath}
\usepackage{bm}
\usepackage{lineno,hyperref}
\usepackage{gensymb}
\usepackage{xcolor}
\usepackage{listings}
\definecolor{codegreen}{rgb}{0,0.6,0}
\definecolor{codegray}{rgb}{0.5,0.5,0.5}
\definecolor{codepurple}{rgb}{0.58,0,0.82}
\definecolor{backcolour}{rgb}{0.95,0.95,0.92}

\modulolinenumbers[5]

\def\doubleunderline#1{\underline{\underline{#1}}}

\journal{Journal of \LaTeX\ Templates}

\begin{document}

\begin{frontmatter}

\title{FSEI-GPU: GPU accelerated simulations of the fluid--structure--electrophysiology interaction \\ in the left heart}

\author[1]{Francesco Viola}
\author[2]{{Vamsi Spandan} } 
\author[3]{Valentina Meschini}
\author[4]{Joshua Romero}
\author[4]{Massimiliano Fatica}
\author[5]{Marco D. de Tullio}
\author[6,7,1]{Roberto Verzicco {\footnote{Corresponding author. E-mail: verzicco@uniroma2.it}}}
\address[1]{Gran Sasso Science Institute, L'Aquila, Italy}
\address[2]{John A. Paulson School of Engineering and Applied Sciences, Harvard University, USA.}
\address[3]{Department of Mathematics, University of Rome Tor Vergata, Rome, Italy}
\address[4]{NVIDIA Corporation, 2788 San Tomas Expressway, Santa Clara, CA 95051, USA}
\address[5]{Department of Mechanics, Mathematics and Management, Politecnico di Bari, Italy}
\address[6]{Department of Industrial Engineering, University of Rome Tor Vergata, Rome, Italy}
\address[7]{Physics of Fluids Group, Max Planck Center Twente for Complex Fluid Dynamics, MESA+ Research Institute, and J. M. Burgers Center for Fluid Dynamics.}

\cortext[mycorrespondingauthor]{Corresponding author}
\ead{verzicco@uniroma2.it}

\begin{abstract}
The reliability of cardiovascular computational models depends on the accurate solution of the hemodynamics, the realistic characterization of the hyperelastic and electric properties of the tissues along with the correct description of their interaction. The resulting fluid--structure--electrophysiology interaction (FSEI) thus requires an immense computational power, usually available in large supercomputing centers, and requires long time to obtain results even if multi--CPU processors are used (MPI acceleration).
In recent years, graphics processing units (GPUs) have emerged as a convenient platform for high performance  computing, as they allow for considerable reductions of the time--to--solution.
\\ \indent
This approach is particularly appealing if the tool has to support medical decisions that require solutions within reduced times and possibly obtained by local computational resources. Accordingly, our multi-physics solver \cite{fsei} has been ported to GPU architectures using CUDA Fortran to tackle fast and accurate hemodynamics simulations of the human heart without resorting to large--scale supercomputers.
This work describes the use of CUDA to accelerate the FSEI on heterogeneous clusters, where both the CPUs and GPUs are used in synergistically with minor modifications of the original source code. The resulting GPU accelerated code solves a single heartbeat within a few hours (from three to ten depending on the grid resolution) running on premises computing facility made of few GPU cards, which can be easily installed in a medical laboratory or in a hospital, thus opening towards a systematic computational fluid dynamics (CFD) aided diagnostic.

\section*{Program summary}
\noindent
\textit{Program Title}: GPU Accelerated Cardiovascular Flow Simulator \\
\textit{Programming language}: Fortran 90, CUDA Fortran, MPI \\
\textit{External routines}: PGI, CUDA Toolkit, FFTW3, HDF5 \\
\textit{Nature of problem}: Fluid--Structure--Electrophysiology interaction (FSEI) in the heart. \\
\textit{Solution method}:  Second--order finite--difference Direct Numerical Simulations of Navier-Stokes equations, 
Immersed Boundary Method for interfacial boundary conditions, Interaction Potential for deforming boundaries, Electrophysiology bidomain model for muscular activation, slab distributed MPI parallelization, GPU accelerated routines.
\end{abstract}

\begin{keyword}
Cardiovascular flows\sep Hemodynamics\sep Multiphysics model \sep Computational engineering
\end{keyword}

\end{frontmatter}


\section{Introduction}
The human heart is a hollow muscular organ that pumps blood throughout the body, to the lungs, and to its own tissue. It drives the systemic--, pulmonary--, and coronary--circulations  to bring oxygen and nutrients to every body cell and to remove the waste products. The heart achieves these fundamental goals by two parallel volumetric pumps, the right and the left, which beat approximately $10^5$ times per day to deliver a continuous flow rate of about 5~l/min with outstanding reliability. This is possible because of the highly cooperative and interconnected dynamics of the heart in which every element is key for the others. In a few words, each heart beat is triggered by specialized pacemaker cells that generate rhythmical electrical impulses propagating along well defined paths and with precise timings thus stimulating a sequence of contractions driving the blood from atria to ventricles and eventually to the arteries. The resulting hemodynamics yields shear stresses and pressure loads on the endocardium and on the valves, whose opening/closing ensures the correct flow direction across heart chambers: only the synchronized and synergistic action of the myocardium electrophysiology, mechanics of the tissues and hemodynamics allows the heart of an adult human to operate on a power of only 8~W, lifelong. 
\\ \indent 
Such a perfect and highly sophisticated mechanism, in which even a minor malfunctioning impairs its pumping efficiency, calls for a complete study on account of the scientific, social, and economic implications.
Concerning the latter we note that cardiovascular disorders (CVD) are the main cause of population death and health care costs of developed countries and, despite the advances of medical research, CVD expenditure projections for the next decades are predicted to become unsustainable. This scenario requires novel approaches that improve the effectiveness of the available diagnostic tools without concurrently increasing  the costs further: computational science can be key for this purpose since it can add predicting capabilities and improve the precision of many of the current evidence based procedures \cite{sackett1997evidence}.
Computer simulations of the blood flow in the heart and arteries can be a precious tool to improve the predicting capabilities of diagnostics, to refine surgical techniques, and to test the performance of prosthetic devices, see figure~\ref{fig:EFScoupling}(a). However, the reliability of cardiovascular simulations depends on the accurate modeling of the hemodynamics, the realistic characterization of the tissues, and the correct description of the fluid--structure--electrophysiology interaction (FSEI) \cite{fsei}.
\\ \indent
Our group has made progress towards the development of a fully--coupled multi--physics computational model for the heart.
In particular, the pulsatile and transitional character of the hemodynamics is obtained by solving directly the incompressible Navier--Stokes equations using a staggered finite--difference method  embedding various immersed boundary (IB) techniques to handle complex moving and deforming geometries. The structural mechanics is based on the interaction potential method \cite{Fedosov,hammer2011mass} to account for the mechanical properties of the biological tissues, which are anisotropic and nonlinear.
The electrophysiology, responsible for the activation potential propagation through the cardiac tissue triggering the active muscular tension, is incorporated by a bidomain model \cite{tung1978} coupled with tenTusscher--Panfilov cellular model \cite{ten2006}. All these models are fully coupled with each other for the resulting computational framework to provide realistic cardiovascular simulations both in terms of muscular activation, intraventricular hemodynamics and wall shear stresses. The three-way FSEI makes the computational model predictive, thus opening the way to numerical experiments for virtually testing new prosthetic devices and surgical procedures.
\\ \indent 
This technological breakthrough, however, is limited by the high computational cost of the multiphysics model where the fluid, structure, and electrophysiology solvers are strongly interconnected, and they should be solved simultaneously in time. On the other hand, 
the time advancement is achieved by discrete time steps whose size is
physically limited by the fastest dynamics (the elastic frequency of the stiff ventricle myocardium) and a tiny time step (in the order of 1$\mu s$) is needed to ensure numerical stability. Such a restriction implies that about half of a million time steps are needed to advance a single heart beat and the computational model has to be highly optimized to resolve a heart beat within few hours in order to timely provide statistically converged results for clinical decision. Efficient code parallelization and effective use of the computational resources are thus essential for clinical application where accurate and timely simulation results are needed.
\\ \indent 
Driven by the above motivations, the FSEI code for cardiac simulations  \cite{fsei} has been ported to GPU architectures as described in this paper. 
The latest GPU technology is indeed well-suited to address those problems, which can be executed on many multi--threaded processors even in double--precision calculations. Furthermore, the high memory bandwidth of recent GPU cards copes well with those algorithms where large arrays need to be stored and modified at any time step. As will be detailed in later sections, our numerical methodology relies on performing calculations on both structured exahedral grids and unstructured triangulated mesh networks. While pure CPU parallelisation has been useful in scaling up such calculations, employing GPU architectures have the potential to provide unprecedented speed-ups with minimal changes to the underlying numerical algorithm and the corresponding code.  
The porting relies on CUDA Fortran \cite{cudabook} that extends Fortran by allowing the programmer to define Fortran functions, called kernels, 
and on the CUF kernel directives that automatically run single and nested  loops on the GPU card without modifying the original CPU code nor needing a dedicated GPU subroutine.
Owing to the enhanced strong scaling properties, the GPU--accelerated FSEI algorithm can now tackle complex cardiac simulations, including the solution of the incompressible Navier--Stokes equations for the hemodynamics -- which is the most demanding solver in terms of computational load -- in a shorter time, thus strongly reducing the time--to--solution to support  medical decision.
\\ \indent 
The paper is organized as follows. In Section~2, the FSEI physical models and solution procedures are reviewed, and in Section 3, the GPU implementation is detailed before discussing the performance of the accelerated code in Section 4. In Section 5, we conclude the paper with a presentation of 
a cardiac simulation of the left human heart. The main conclusions and perspectives for future developments are in given in Section 6.

\section{The Fluid--Structure--Electrophysiology interaction (FSEI) }\label{sec:fff} 
 \begin{figure}[t!]
\centering
\includegraphics[width=.95\textwidth]{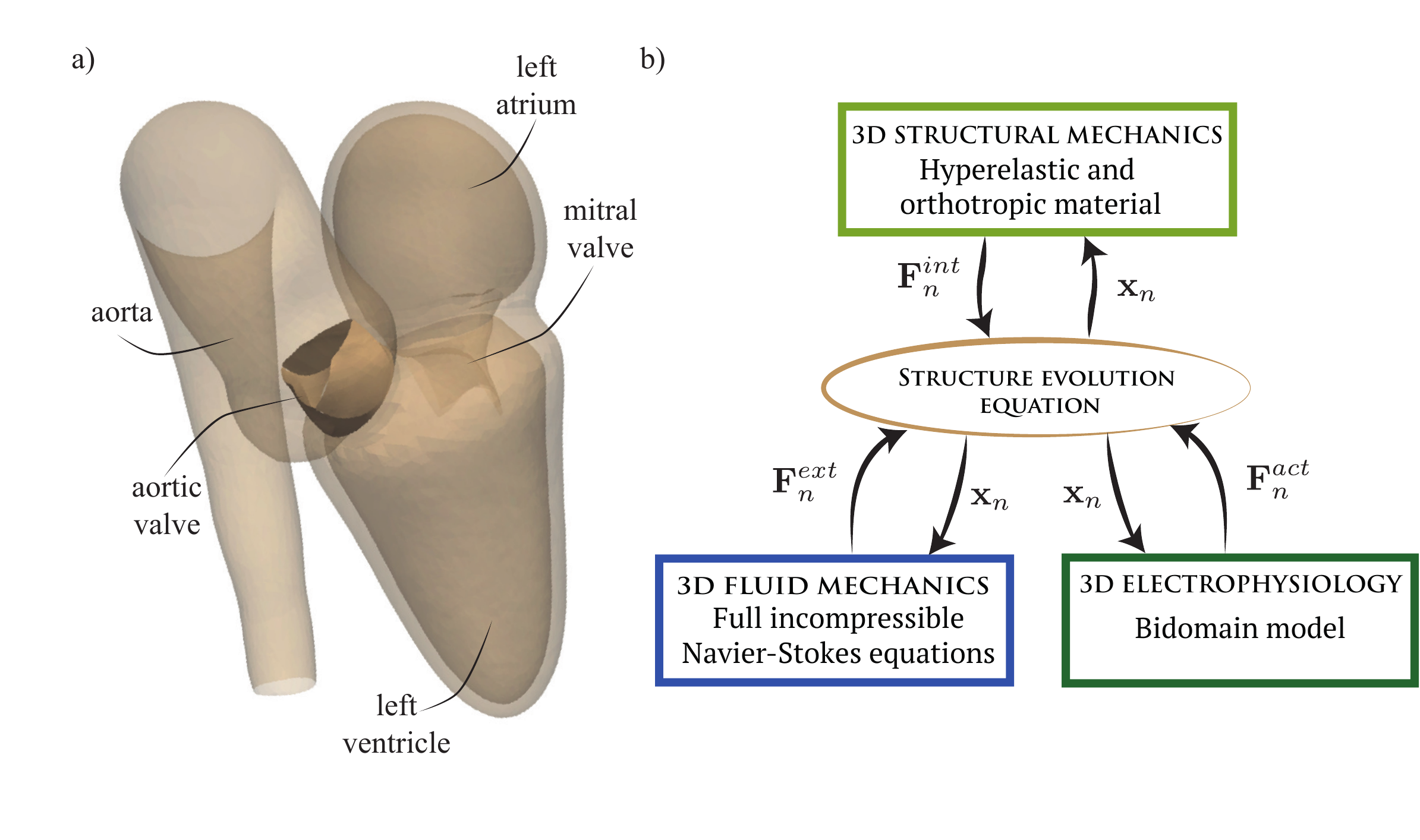}
\caption{a) Computational domain of the human left heart. b) Sketch of the fluid--structure--electrophysiology (FSEI) coupling. }
\label{fig:EFScoupling}
\end{figure}
In this section, the fluid, the structural, and the electrophysiology solvers  along with their coupling strategy are briefly introduced.
A typical cardiac geometry is shown in Figure~\ref{fig:EFScoupling}(a), where the myocardium of the left heart chambers is discretized using an unstructured tetrahedral mesh in the form of a VTK or Gmsh file containing the information about the spatial positions of the vertices of the tetrahedral cells. On the other hand, the geometry of slender structures such as the valve leaflets and the arteries is provided as a triangulated surface using the GTS (GNU Triangulated Surface) format listing the node positions, the index of the nodes connected by an edge and the index of the edges belonging to the same triangular face.
\\ \indent 
As sketched the diagram of Figure~\ref{fig:EFScoupling}(b), the contraction and relaxation of the heart chambers along with the aorta and valve leaflets kinematics results from the dynamic balance between inertia, external $\mathbf F_n^{ext}$, passive~$\mathbf F_n^{int}$, and active~$\mathbf F_n^{act}$ forces acting on each mesh node. The  Newton's second law of motion yields
\begin{equation}\label{eq:newton}
m_n\frac{ \mathrm{d}^2 \mathbf x_n}{\mathrm{d} t^2} = \mathbf F_n^{ext} + \mathbf F_n^{int} +  \mathbf F_n^{act}, 
\end{equation}
where  $\mathbf x_n$ is the (instantaneous) node position and $m_n$ its mass (see Section~\ref{sec:structure}).
The hydrodynamic force is non--zero only on the mesh nodes placed on the wet surfaces (e.g. the endocardium in the heart chambers, the inner wall of the aorta, and the valve leaflets), whereas the active tension can be non--zero only for the nodes belonging to the muscular myocardium, i.e. ventricles and atria.
\\ \indent  
In principle, all the forces at the right--hand--side of Equation~\eqref{eq:newton} should be calculated simultaneously since they are all function of the unknown instantaneous geometry of the tissues and vice--versa, thus calling for an iterative approach. We have implemented both strong and loose coupling procedures in the code. The first is based on a  predictor--corrector two--step Adams--Bashforth scheme and the three solvers are iterated (typically $2$--$3$ times) until the maximum relative error of the nodes position and velocity drops below a prescribed threshold (equal to $10^{-7}$ for nondimensional quantities). Conversely, in the loose coupling, the blood flow and the electrophysiology are solved first and the generated hydrodynamic and active loads are used to evolve the structure according to Equation~\eqref{eq:newton}. 
Dedicated numerical tests showed that, since the time step size is constrained by the elastic stiffness of the myocardium, the loose coupling approach is seen to be stable and yields an overall lower computational cost with respect to the strong coupling, while retaining the same accuracy and precision. We refer to \cite{Meschini2018,fsei,fseichap} for a comprehensive discussion and numerical tests.
\\ \indent 
In the following sections we briefly review the fluid, structural and electrophysiology solvers providing the forces governing the heart tissues kinematics, namely $\mathbf F_n^{ext},~\mathbf F_n^{int}~\text{, and}~\mathbf F_n^{act}$.

\subsection{Fluid and pressure solver}\label{sec:NS}
\begin{figure}[t!]
\centering
\includegraphics[width=1\textwidth]{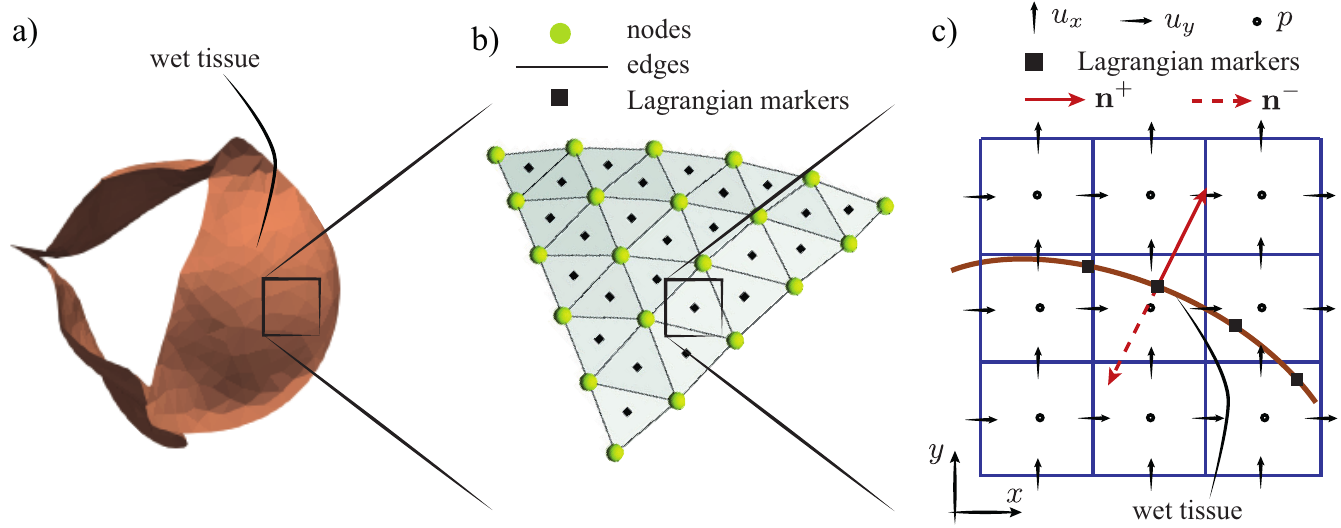}
\caption{IB treatment of the deformable tissues. (a) Generic wet surface, (b) triangulated mesh with the mass concentrated at the nodes and the Lagrangian markers placed at its centroids, (c) support domain around a Lagrangian marker  consisting of 27 Eulerian cells.}
\label{fig:mls}
\end{figure}

The hematic velocity $\mathbf u$ and pressure $p$  are governed by the incompressible Navier--Stokes and continuity equations which in non--dimensional form read:
\begin{equation}\label{NS}
\begin{split}
 \frac{\partial {\bf u}}{\partial t} +  \nabla \cdot  ( {\bf u {\bf u}}) &=  - \nabla p + \nabla \cdot { \doubleunderline{ \tau} }  +{\bf f},  \\
\nabla \cdot {\bf u} &= 0,
\end{split}
\end{equation}
with $\doubleunderline{ \tau }$ the viscous stress tensor, which depends on the strain rate tensor $\doubleunderline{E} =0.5(\nabla \mathbf u + \nabla^T \mathbf u)$ according to the Carreau--Yasuda blood model (shear--thinning) as detailed in \cite{Katritsis2007,DeVita}. 
In the case of hematic flows in the heart chambers and/or main vessels, however, the blood can be modelled as a Newtonian fluid (by changing a flag in the code) with the linear constitutive relation $\doubleunderline{\tau} = 2 Re^{-1} \doubleunderline{E}  $ as the non--Newtonian fluid features manifest only in vessels of sub--millimeter diameter. 
\\ \indent 
The governing equations~\eqref{NS} are solved over Cartesian meshes using the AFiD solver, based on central second--order finite--differences discretized on a staggered mesh \cite{raimoin,Verzicco1996,pencil}, and the no--slip condition on the wet heart tissues is imposed using an IB technique based on the moving least square (MLS) approach \cite{uhlmann2005immersed,Vanella2009,TuPa}.
The first of equations~\eqref{NS} is discretized in time using an explicit Adams--Bashforth method for the nonlinear convective term and an implicit Crank-Nicolson method for the viscous terms:
\begin{equation}\label{eq:NSdiscr0}
\frac{ \mathbf u^{n+1} - \mathbf u^{n}}{\Delta t} +  [\gamma \nabla \cdot  ( {\bf u {\bf u}})^n + \rho \nabla \cdot  ( {\bf u {\bf u}})^{n-1} ] = -\nabla p^{n+1} + \frac{1}{2 Re} \nabla^2 ( \mathbf u^{n+1} + \mathbf u^n) + \mathbf f,
\end{equation}
with the superscripts $n$ and $n+1$ indicating the velocity and pressure fields at time $t^n$ and $t^{n+1}=t^n +\Delta t$, with $\Delta t$ the time step.
In incompressible flows, the instantaneous pressure field $p^{n+1}$ does not have a dynamic role, but it acts only as a Lagrangian multiplier  assuring the solenoidal condition for the velocity field $\mathbf u^{n+1}$ imposed by mass conservation. For this reason, only the updated pressure field $p^{n+1}$ is used in \eqref{eq:NSdiscr0}, rather than a time average between the time levels $n$ and $n+1$.
The numerical coefficients $\gamma$ and $\rho$ appearing in Equation~\eqref{eq:NSdiscr0} depend on the temporal integration schemes of the convective terms and are equal to $3/2$ and $-1/2$, respectively, for the Adams--Bashforth scheme (although not  reported here for the sake of conciseness, a third order Runge--Kutta scheme is also implemented in the code).
Since is not possible to solve simultaneously Equation~\eqref{eq:NSdiscr0} for $\mathbf u^{n+1}$ and $p^{n+1}$, a fractional--step method \cite{kim1985application,Verzicco1996} is used
and the no-slip boundary condition is then imposed on some Lagrangian markers uniformly distributed on the immersed boundary domain and then transferred to several Eulerian grid-points as shown in Figure~\ref{fig:mls}.
A three--dimensional support domain consisting of $N_e = n \times n \times n $ Eulerian nodes ($n=3$ is typically  used)  is created around  each Lagrangian marker, and the fluid velocity at the body position $\mathbf u^n(\mathbf x_b)$ is computed interpolating the velocity of the $N_e$ Eulerian grid-points in the support domain as
\begin{equation}\label{eq:interpv}
u_i(\mathbf x_b)  = \sum_{k=1}^{N_e}  \phi^k_i (\mathbf x_b) u_i(\mathbf x_k) ,
\end{equation}
where the $\phi^k_i (\mathbf x)$ are the transfer operators which depend on the shape functions used for the interpolation.  In this paper, a linear basis function is used, $\mathbf p^T(\mathbf x)=[1,x,y,z]$, with an exponential weight function centered at the location of the Lagrangian marker \cite{TuPa}.
The interpolated velocity~\eqref{eq:interpv} is used to compute the IB force at the exact location of the marker which is then transferred back to the Eulerian grid-points as a distributed forcing.
This procedure is applied to all Lagrangian markers for the three velocity components, and the resulting IB forcing is applied to update the intermediate velocity.
\\ \indent In order to provide the hydrodynamic loads as input to the structural solver for fluid--structure coupling, the pressure and the viscous stresses are evaluated at the Lagrangian markers laying on the immersed body surface.
In the case of the valve leaflets, both sides of the tissues are wet by the hematic flow and the local hydrodynamic force at the wet triangular face $\mathbf F_f^{ext} $ is computed along both the positive $\mathbf n^+$ and negative $\mathbf n^-=-\mathbf n^+$ normal directions: 
\begin{equation}\label{eq:Fext2}
\mathbf F_f^{ext} =  [ -(p^+_f - p^-_f) \mathbf n^+_f  + \bm ( \doubleunderline{\tau}^+_f - \doubleunderline{\tau}^-_f) \cdot \mathbf  n^+_f ] A_f,
\end{equation}
where $A_f$ is the area of the triangular face.
On the other hand, for singe--side wet surfaces, like the ventricle, aorta and atrium, hydrodynamic loads are only computed over the inner surface.
\begin{equation}\label{eq:Fext}
\mathbf F_f^{ext} =  [ -p_f \mathbf n_f  + \doubleunderline{\tau}_f \cdot \mathbf  n_f] A_f,
\end{equation}
where $\mathbf n_f$ is the normal vector pointing towards the hematic flow wetting the surface.
The hydrodynamic loads evaluated at the wet faces are then transferred to the wet nodes as follows
\begin{equation}\label{eq:Fextn}
\mathbf F_n^{ext} =\frac{1}{3}  \sum_{i=1}^{N_{nf}}  \mathbf F_{fi}^{ext} A_{fi},
\end{equation}
where $N_{nf}$ is the number of faces sharing the node $n$, $ \mathbf F_{fi}^{ext} $ and $A_{fi}$ are the hydrodynamics force  and surface of the $i$--th face sharing the node $n$. 

\subsection{Structural mechanics} \label{sec:structure}
The dynamics of the deformable heart tissues is solved using a spring--network structural model based on an interaction potential approach  \cite{Fedosov, hammer2011mass, TuPa}. A three--dimensional (3D) solver is used for the ventricular and atrial myocardium, whereas a two--dimensional (2D) one is adopted for thin membranes as the valve leaflets and the aorta.
\begin{figure}[t!]
\centering
\includegraphics[width=.95\textwidth]{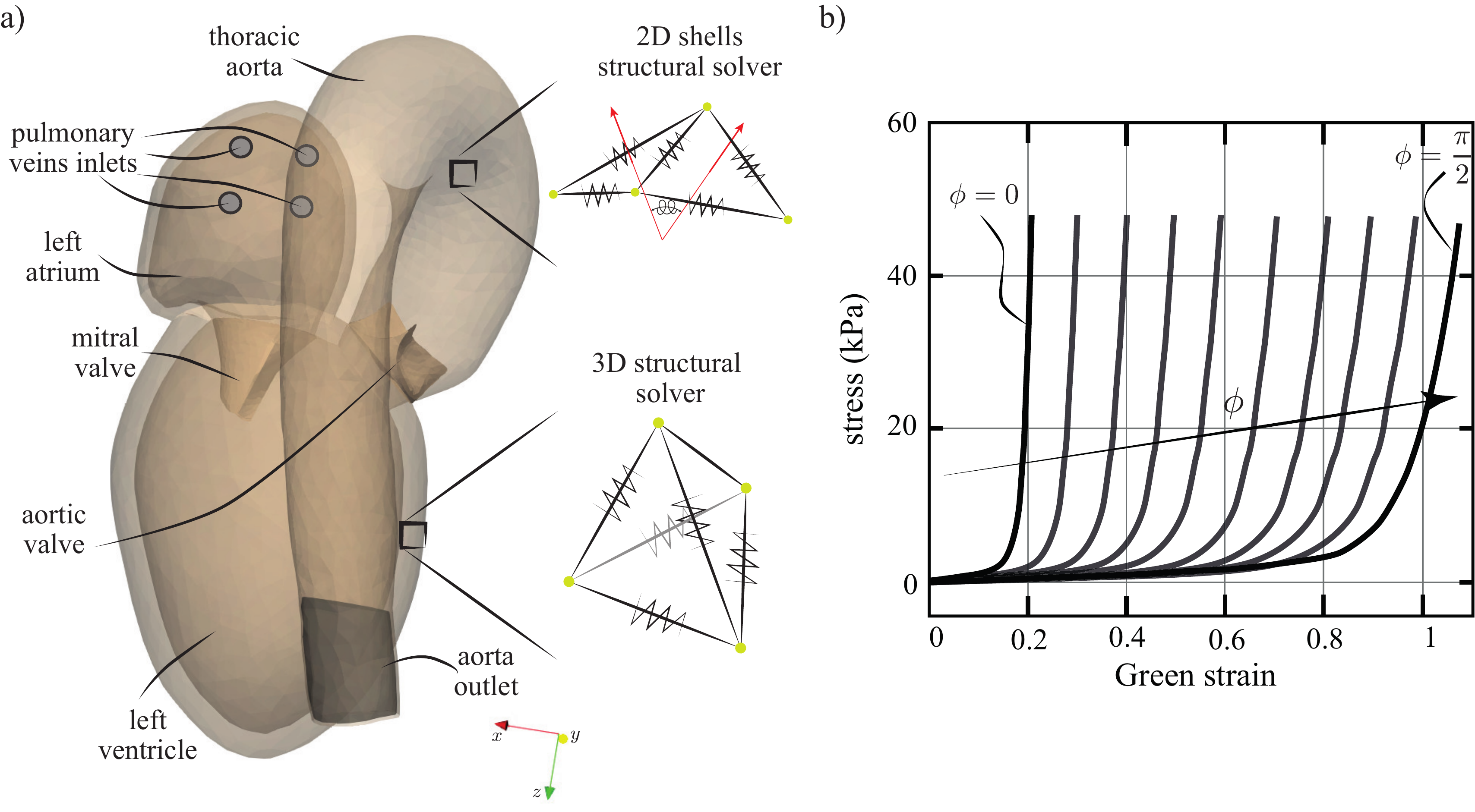}
\caption{(a) Sketch of the left cardiac configuration. The 3D myocardium is discretized using a tetrahedral mesh, while the 2D valve leaflets and arteries are discretized using triangular elements. A nonlinear spring is placed at each mesh edge (b) with a  hyperelastic and anisotropic constitutive relation depending on the spring orientation with respect to the local fiber direction, $\phi$. In the case of 2D structures their bending stiffness is obtained by placing some out-of-plane springs connecting the centroids of two adjacent triangular faces, see first inset in (a). }
\label{fig:vangelder}
\end{figure}
\\ \indent 
The 3D structural model is built considering a tetrahedral discretization of the ventricular and atrial  myocardium (same grid used by the electrophysiology solver) and placing a spring on each edge of the network, which yields a 3D force field as a response to stretching as shown in the lower inset of figure~\ref{fig:vangelder}(a). On the other hand, the 2D structural model for the cardiac valve leaflets and vessels is based on surface triangular meshes as indicated in the upper inset of the same panel. Although the 3D and 2D spring models were proposed in the framework of linear elastic materials \cite{vangelder}, they  have been extended to the case of hyperelastic and anisotropic materials so to correctly model the biological cardiac tissues, in a similar fashion to the method proposed by \cite{hammer2011mass,TuPa,fsei} for 2D shells.
At any point of the myocardium, the elastic stiffness is, indeed, larger in the fiber direction, $\hat{\mathbf e}_f$ than in the sheet $\hat{\mathbf e}_s$ and sheet--normal $\hat{\mathbf e}_n$ directions (anisotropic behaviour) and increases nonlinearly with the strain (hyperelastic behaviour).  
According to a Fung--type constitutive relation, the strain energy density reads:
\begin{equation}\label{eq:potentiallinear}
W_e = \frac{c}{2}(e^Q-1),
\end{equation}
with  $Q=\alpha_f \epsilon_{\mathit{ff}}^2+\alpha_s \epsilon_{\mathit{ss}}^2 +\alpha_n \epsilon_{\mathit{nn}}^2 $ being a combination of the Green strain tensor components \cite{hammer2011mass} in the fiber, $\epsilon_{\mathit{ff}}$, sheet, $\epsilon_{\mathit{ss}}$, and sheet--normal $\epsilon_{\mathit{nn}}$ directions. The general expression for $Q$ \cite{hammer2011mass,TuPa}, which includes also the cross terms of the Green strain tensor, has been simplified under the assumption of pure axial loading and, consequently, the non--null second Piola--Kirchhoff stress tensor components in the three direction read $\tau_{\mathit{ff}}=c \alpha_{\mathit{ff}} e^{\alpha_{\mathit{ff}} \epsilon^2_{\mathit{ff}}} \epsilon_{\mathit{ff}}$, $\tau_{\mathit{ss}}=c \alpha_{\mathit{ss}} e^{\alpha_{\mathit{ss}}\epsilon^2_{\mathit{ss}}} \epsilon_{\mathit{ss}}$ and 
$\tau_{\mathit{nn}}=c \alpha_{\mathit{nn}} e^{\alpha_{\mathit{nn}}\epsilon^2_{\mathit{nn}}} \epsilon_{\mathit{nn}}$. 
The latter two terms can be taken as equal since it is found experimentally that $\alpha_{\mathit{nn}}=\alpha_{\mathit{ss}}$ \cite{costa1996three,usyk2000effect}, meaning that the local axial stress of the mesh springs only depends on their inclinations, $\phi$, with respect to the local fiber direction.
Hence, the local stress within an edge inclined by $\phi$ with respect to the local fiber direction is computed as  
\begin{equation} \label{eq:stress}
\tau_{\phi} = c \alpha_{\phi} e^{\alpha_{\phi}\epsilon_\phi^2} \epsilon_\phi,
\end{equation}
where $\alpha_\phi=\sqrt{\alpha_\mathit{ff}^2 \cos^2 \phi+\alpha_\mathit{nn}^2 \sin^2 \phi}$ (we recall, assuming $\alpha_{\mathit{nn}}=\alpha_{\mathit{ss}}$),  and the strain $\epsilon_\phi$ is calculated as the spring elongation relative to its instantaneous length, i.e. $\epsilon_\phi = (l-l_0)/l$,
being $l$ and $l_0$, the actual and the stress--free length of the edge, respectively. 
As indicated in figure~\ref{fig:vangelder}(b), the stress depends linearly on the strain for small strain values and grows exponentially for larger ones, on the other hand the stiffness is inversely correlated with the angle $\phi$.
The corresponding force in 3D tissues applied to the nodes~$n_1$~and~$n_2$ sharing the edge $l_{n_1,n_2}$ thus reads:
\begin{equation}\label{eq:vg3}
\begin{split}
 \mathbf F^{int \text{3D}}_{n_1} =  \underbrace{ \tau_\phi}_{\text{stress}} \underbrace{ \sum_{j=1}^{N_{n_1,n_2}} \frac{V_{cj}}{l_{n_1,n_2}}  }_{\text{tissue cross--section}} \underbrace{ \frac{ \mathbf x_{n_1} - \mathbf x_{n_2}}{l_{n_1,n_2}}}_{\text{force direction}},\hspace{1.cm}  \mathbf F^{el}_{n_1} = -\mathbf F^{el}_{n_2},
\end{split}
\end{equation}
with $\mathbf x_{n_1}$ ($\mathbf x_{n_2}$) the position of the node~$n_1$ ($n_2$) and $V_{cj}$ the area of the $j-th$ tetrahedron out of  the $N_{n_1,n_2}$ ones sharing the edge $l_{n_1,n_2}$.  
\\ \indent
On the other hand, the nonlinear elastic force in 2D tissues applied to a couple of adjacent nodes sharing an edge reads
\begin{equation} \label{eq:vg2}
\begin{split}
 \mathbf F^{el \text{2D}}_{n_1} = \underbrace{\tau_\phi}_{\text{stress}} \underbrace{  s\frac{
 A^{(1)}_{n_1,n_2}+A^{(2)}_{n_1,n_2}
}{l_{n_1,n_2}} }_{\text{tissue cross--section}} \underbrace{ \frac{ \mathbf x_{n_1} - \mathbf x_{n_2}}{l_{n_1,n_2}}}_{\text{force direction}},\hspace{1.cm}  \mathbf F^{el}_{n_2} = -\mathbf F^{el}_{n_1},
\end{split}
\end{equation}
with $\mathbf x_{n_1}$ ($\mathbf x_{n_2}$) the position of the node~$n_1$ ($n_2$) and $A^{(1,2)}_{n_1,n_2}$ is the area of the two triangles sharing the edge $l_{n_1,n_2}$. The parameters of the Fung constitutive relation can be set so as to reproduce the stress-strain curves in the fiber and cross-fiber direction measured in the ex-vivo experiments \cite{fsei,fseichap}.
\\ \indent 
Since in the 2D spring--network the axial loading~\eqref{eq:vg2} only accounts for the in-plane stiffness, an additional bending energy term has to be included so that to provide the out-of plane bending stiffness to the shells.
The out--of--plane deformation of two adjacent triangles sharing an edge is then associated with an elastic energy due to the contraction/expansion of a bending spring, whose energy involves four adjacent nodes as shown in the right inset of figure~\ref{fig:vangelder}(a).
Considering two adjacent triangular faces sharing an edge that are inclined of an angle $\theta$, the discretized bending energy is equal to \cite{kantor1987}:
\begin{equation}\label{eq:bendpot}
W_b =k_b [ 1- \cos(\theta - \theta_0)],
\end{equation}
where $\theta_0$ is the initial inclination of the stress--free configuration.
The bending constant is equal to $k_b = 2B/\sqrt{3}$ \cite{li2005,TuPa}, with $B=c \alpha_\phi s^3/[12(1-\nu^2_m)]$ the bending modulus of a planar structure, where $s$ is the tissue thickness, $c \alpha_\phi$ is the equivalent Young modulus in the limit of small strain (that depend on the Fung properties of the tissues) and $\nu_m=0.5$ is the Poisson ratio of the material. The corresponding bending nodal forces, $\mathbf F_n^{be 2D}$ can be then obtained by taking the gradient of the bending potential~\eqref{eq:bendpot} as detailed in \cite{TuPa} and the passive internal forces of shell structures at a given node thus read 
$ \mathbf F_n^{int 2D} =\mathbf F_n^{el 2D} + \mathbf F_n^{be 2D}$.
\\ \indent 
In the 3D (2D) structural models, the mass of the tissue is concentrated on the mesh nodes proportionally to the volume of the tetrahedrons (area of the triangles) sharing a given node. In the case of a 3D (2D) tissue of local density $\rho_{cj}$ ($\rho_{fj}$) the mass of  the $j$--th cell (face) with volume $V_{cj}$ (surface $A_{fj}$) is equally distributed among its four (three) nodes and the mass of a node, $m_n$, reads
 \begin{equation}\label{eq:mass3}
 m_n^{\text{3D}} = \frac{1}{4} \sum_{j=1}^{N_{nc}} \rho_{cj}  V_{cj}, ~~~ \left( m_n^{\text{2D}} = \frac{1}{3} \sum_{j=1}^{N_{nf}} \rho_{fj} s_{fj} A_{fj}   \right),
 \end{equation}
being the summation extended only to the  $N_{nc}$ tetrahedrons ($N_{nf}$ triangles) sharing the selected node $n$ and $s_{fj}$ the local thickness of the deformable shell.

\subsection{Electrophysiology}\label{sec:electro}
The electrical activation of the myocardium is governed by the bidomain model, called in this way because of the conductive media modelled as an intracellular and an extracellular overlapping continuum domains separated by the myocytes membrane \cite{tung1978,Clayton2008}. 
The potential difference across the membrane of the myocytes, the transmembrane potential $v$ and the extracellular potential $v_{ext}$  satisfy:
\begin{equation}\label{eq:electro}
\begin{split}
  \chi \left(   C_m \frac{\partial v}{\partial t}  +I_{ion}(\eta) + I_s   \right)  &= \nabla \cdot (\doubleunderline{M}^{int} \nabla v) + \nabla \cdot (\doubleunderline{M}^{int} \nabla v_{ext}) ,  \\
    0  &=             \nabla \cdot  (\doubleunderline{M}^{int} \nabla v + (\doubleunderline{M}^{int}+\doubleunderline{M}^{ext}) \nabla v_{ext})),            \\
    \frac{\partial \eta}{\partial t}& = F(\eta,v,t)
\end{split}
\end{equation}
where $\chi$ and $C_m$ are the surface--to--volume ratio of cells and the membrane capacitance $C_m$, respectively.
The parameters $\doubleunderline{M}^{int} $ and $\doubleunderline{M}^{ext}$ are the conductivity tensors of the intracellular and extracellular media, which reflect the orthotropic myocardium electrical properties and depend on the local 
fiber orientation, with the electrical signal propagating faster along the muscle fiber than in the cross--fibers 
directions. The conductivity tensor in the global coordinate system are thus obtained by the transformations
$\doubleunderline{M}^{ext} =\doubleunderline{\mathcal A} \hat{\doubleunderline{M}}^{ext} \doubleunderline{\mathcal A}^T$ and 
$\doubleunderline{M}^{int} = \doubleunderline{\mathcal A} \hat{\doubleunderline{M}}^{int} \doubleunderline{\mathcal A}^T$,
where $\doubleunderline{\mathcal A}$ is the rotation matrix containing column--wise the components of fiber, sheet and sheet--normal unit vectors 
and $\doubleunderline{\hat{M}}^{ext}, ~\doubleunderline{\hat{M}}^{int}$ are diagonal  tensors expressed in the principal basis formed by the fiber, sheet and sheet--normal directions, where its non--null diagonal components are the principal electrical conductivities~\cite{sundnes2007}. 
\\ \indent The set of equations~\eqref{eq:electro} are discretized on the same tetrahedral mesh used for the three--dimensional structural solver by using an in--house finite volume (FV) library, which provides a suitable approach for solving the electrophysiology equation in complex geometries \cite{fseichap}. The FV method is \textit{cell--based} \cite{moukalled2016finite} meaning that the unknown fields are defined at the center of each cell and, using the divergence theorem, the bidomain equations~\eqref{eq:electro} can be written in conservative form on each tetrahedron, $\Omega_i$. Furthermore, assuming all quantities to be uniform on the faces of the tetrahedrons (as typically done in FV) and adopting an explicit time scheme, the first equation of the system~\eqref{eq:electro} can be solved over each tetrahedron face:
\begin{equation}\label{eq:electro2}
\begin{split}
&   C_m \frac{ v_i^{n+1}-v_i^{n}}{\Delta t}   =\frac{\gamma_2}{  \chi  V^{n}_{\Omega_i} } \sum_{j=1}^4 A^n_{\partial \Omega_{i,j}}   [ \doubleunderline{M}_i^{int} (\nabla v_i^n + \nabla v_{ext~i}^n ) ]_j \cdot \mathbf n_j    \\
&+  \frac{\rho_2}{  \chi  V^{n-1}_{\Omega_i} } \sum_{j=1}^4 A^{n-1}_{\partial \Omega_{i,j}}   [ \doubleunderline{M}_i^{int} (\nabla v_i^{n-1} + \nabla v_{ext~i}^{n-1} )  ) ]_j \cdot \mathbf n_j \\
& -\gamma_2(I^n_{ion,i} + I^n_{s,i} )  -\rho_2(I^{n-1}_{ion,i} + I^{n-1}_{s,i} ),  
\end{split}
\end{equation}
where $v_i^{n-1}$, $v_i^{n}$ and $v_i^{n+1}$ are the transmembrane potentials defined at the $i$-th cell having a volume of  $V_{\Omega_i}$  at the time $t^{n-1}=t^{n}-\Delta t$, $t^{n}$ and $t^{n+1}=t^{n}+\Delta t$. 
The gradients $\nabla v_i^n$ and $\nabla v_i^{n-1}$ (as well as $\nabla v_{ext~i}^n$ and $\nabla v_{ext~i}^{n-1}$) are defined at the face cell and are obtained by interpolating the gradients at the two cells sharing the face, which have been obtained using the Gauss-Green formula 
\begin{equation}\label{eq:electro3}
\nabla v_c  = \frac{1}{V_c} \sum_{j=1}^4 v_{fj} S_{fj} \mathbf n_{fj}, 
\end{equation}
where the subscripts $c$ and $f$ indicate quantities evaluated at the mesh cells and faces, and the summation index $j$ loops over the four faces of the tetrahedral cell having surfaces $A_{\partial \Omega_i}$. 
Once $v^{n+1}$ is solved through equation~\eqref{eq:electro2}, the external potential at time $t^{n+1}$, $v_{ext~i}^{n+1}$, is obtained by solving the linear system given by the second equation of the system~\eqref{eq:electro} using an iterative GMRES method with restart \cite{trefethen1997numerical}.
The time--scheme coefficients are respectively equal to $\gamma=1,~\rho=0$ for first--order backward Euler method and to $\gamma=3/2,~\rho=-1/2$ for second--order Adam-Bashfort methods.
Hence, at each time step the updated transmembrane potential $v^{n+1}$ is obtained as a function of $v$, $v_{ext}$, $I_{ion,i}$ and $I_{s,i}$ evaluated at $t^{n}$ and $t^{n-1}$.
The updated state vector of the cellular model $\eta^{n+1}$ determining the updated ionic current $I_{ion}^{n+1}$ is computed solving a system of 19 coupled nonlinear ODEs of the tenTusscher--Panfilov model on each tetrahedron, which are indicated in compact form by the last equation of the system~\eqref{eq:electro}. 
These equations are known to be stiff, and explicit time schemes generally require prohibitively small time steps to be numerically stable. In contrast, implicit schemes are more stable but also computationally expensive.
This impasse is promptly solved by using the Rush--Larsen method \cite{rush1978practical,marsh2012secrets} where the quasi-linear (gating) variables are solved analytically within a time step if the transmembrane potential $v$ is held constant and an explicit method is used to integrate the remaining nonlinear ones.
\\ \indent
The active muscular tension at each grid node  $\mathbf F_n^{act}$ is then obtained as a function of the transmembrane potential $v$ through the model equation \cite{nash2004} as detailed in \cite{fseichap}.

\section{Code parallelization and GPU acceleration}
In this section, the FSEI parallelization and its GPU acceleration is described. The GPU porting is based on CUDA Fortran \cite{cudabook} as the CPU code was originally written in Fortran90, and the resulting CUDA version keeps the structure of the original CPU Fortran although it allows portions of the computation to be off-loaded to the GPUs.
In CUDA, the code instructions running on the GPU cards are programmed in the \textit{kernel} which is a subroutine launched with a grid of threads grouped into thread blocks. Each thread block runs independently from the others on an available multiprocessor of the GPU, and the thread block data can be shared among threads belonging to the same block.
Importantly, the GPU--accelerated FSEI code not only uses dedicated GPU subroutines but it also makes extensive use of the CUF kernels, which are particularly convenient for porting to GPU single and nested do--loops without modifying its content and simply calling the CUDA directive. The latter appears as a comment to the compiler if GPU code generation is disabled (similar to the OpenMP directives that are ignored if OpenMP is not enabled).  Therefore, when possible, CUF directives allow for a very efficient and easy to implement GPU parallelization \cite{afidgpu}. CUDA--enabled GPUs thus provide thousands of processor cores which allow to run tens of thousands of threads concurrently resulting in an effective speed--up of algebraic operations over large computational grids as is the present case.
\\ \indent 
In the final version of the code the whole fluid, structural and electrophysiology computations are performed on the GPUs, whereas the CPUs are only used for I/O and to stage the data needed during the communication phases. Nevertheless, when the GPU code is compiled omitting the CUDA flags the original CPU code is retrieved.

\subsection{Fluid and pressure solver}
The parallelization of the Navier--Stokes solver introduced in section~\eqref{sec:NS} is based on a domain decomposition where the Cartesian domain is split into slabs \cite{vamsi2017,vamsi2018}. According to this 'one-dimensional slab' parallelisation, each processor needs to store information from the neighbouring processors which is required for computing the derivatives in what is called  a 'halo/ghost' layer and since the flow solver employs a second-order finite difference spatial discretisation at most one halo layer is required on each side of a slab. The viscous terms are treated implicitly yielding the solution of a large sparse matrix, which is avoided by an approximate factorization yielding tridiagonal matrices (one for each direction \cite{Verzicco1996}) inverted using Thomas' algorithm with a Sherman--Morrison perturbation in the two periodic dimensions. 

\subsection{IB-MLS }
The parallelization of the IB method is carried out as in \cite{vamsi2017,vamsi2018}. The wet surfaces of the cardiac valve leaflets, arteries and the endocardium are organized as a whole wet surface whose information (nodes, edges and triangles) is stored in all the processors, although the computations required for each Lagrangian node/structure is performed only by the specific processor, depending on the task that needs to be performed (task--based parallelism).
\\ \indent 
First, all processors determine the three indices of the Eulerian mesh cell containing each marker (centroid) of the Lagrangian mesh, compute the geometrical properties of the triangular face (e.g. area and normal vector) and store this information into global arrays so that it is available to every processor.
The IB forcing $\mathbf f$ applied at the fractional step is then computed by interpolating the flow velocity on the centroids of the Lagrangian mesh using a MLS method~\eqref{eq:interpv} and  each processor performs all the operations required on its respective slabs, hence, the MLS interpolation along with determining the IBM force is performed only on the Lagrangian markers residing within the processors slab regardless of which immersed body it belongs to. 
\\ \indent 
A similar parallelization strategy is used to compute the external forces $\mathbf F_f^{ext}$ on the triangular faces of the wet surfaces (see equations~\eqref{eq:Fext2}~and~~\eqref{eq:Fext}), which are then transferred to the wet nodes according to equation~\eqref{eq:Fextn} and the resulting  $\mathbf F_n^{ext}$ are then communicated over all processes using MPI\_ALLREDUCE.
Both the calculation of the IB forcing $\mathbf f$ and the external forces $\mathbf F_n^{ext}$ have been accelerated by coding dedicated GPU subroutines that are executed in place of the original CPU subroutines when the code is compiled with the -DUSE\_CUDA  flag. Although we use CUF kernel extensively, these subroutines are coded manually on the GPU since the 4x4 system of equations that has to be solved for each Lagrangian marker to compute the $\phi^k_i (\mathbf x_b)$ weights is better handled using multiple threads (namely 16) concurrently.

\subsection{Structural solver}
Through the simulation the instantaneous configuration $\mathbf x_n$ of both the 3D and the 2D structures are organized as a whole body: one for the three-dimensional myocardium of the heart chambers and another comprising the two--dimensional structures such as the cardiac valve leaflets and the arteries. The GPU acceleration of the internal stresses computation corresponding to equations~\eqref{eq:vg2}~and~\eqref{eq:vg3} in section~\ref{sec:structure} is achieved by using the CUF kernel directives that are very simple to use as the original Fortran code is basically unaltered and the GPU acceleration is obtained by computing all the internal forces at the mesh nodes $\mathbf F_n$ simultaneously. On the other hand, the bending forces used for 2D shells only, see equation~\eqref{eq:bendpot}, are computed using a dedicated GPU subroutine since multiple threads may write concurrently on the same array element and the build--in function ATOMICADD has to be used. 
In a similar fashion to the subroutine for the IB forcing, the GPU subroutine are only executed when the CUDA flag is active, the corresponding CPU routine is executed otherwise.
 Note that both the internal and the active (see next section) forces do not depend on the velocity field defined on the Eulerian mesh and, therefore, the computing load is distributed evenly across all processors.

\subsection{Electrophysiology solver}
Owing to the combination of the finite volume formulation that has a diagonal mass matrix and an explicit temporal scheme, the bidomain equations are marched in time 
and the resulting algebraic problem can be efficiently accelerated through few CUF kernels. Specifically, the electrophysiology solver results in a sequence of loops on the mesh cells and on the mesh faces, which are GPU accelerated simply wrapping the original CPU code with CUF kernel directives. 
As an example, at each time step, the gradient of the transmembrane potential can be evaluated in parallel by using the Gauss--Green formula~\eqref{eq:electro3} simultaneously on the mesh cells.
Moreover, the interpolation needed to evaluate the transmembrane potential at the tetrahedral nodes and faces and the gradient at the mesh faces are also parallelized with the same simple approach. 
\\ \indent
The cellular model reproducing the fluxes through the ionic channels and coupled to the bidomain equations as in Equation~\eqref{eq:electro}, calls for the solution of a system of nonlinear ODEs at each mesh cell at any time step.
As the ODEs depend, by definition, only  on the transmembrane potential at the previous time steps rather than on its field variations,
the 19 ODEs of the tenTusscher~Panfilov model over each cell are time marched concurrently by the GPU threads invoked by the CUF kernels.
\\ \indent
The total force acting on each mesh node is computed as the summation of the external forces (pressure + viscous), the internal forces arising from the elastic potentials and the active forces. The Newton equation~\eqref{eq:newton} at each cell node is also solved using a CUF kernel directive.

\section{Code performance}
\begin{figure}[h!]
\centering
\includegraphics[width=0.99\textwidth]{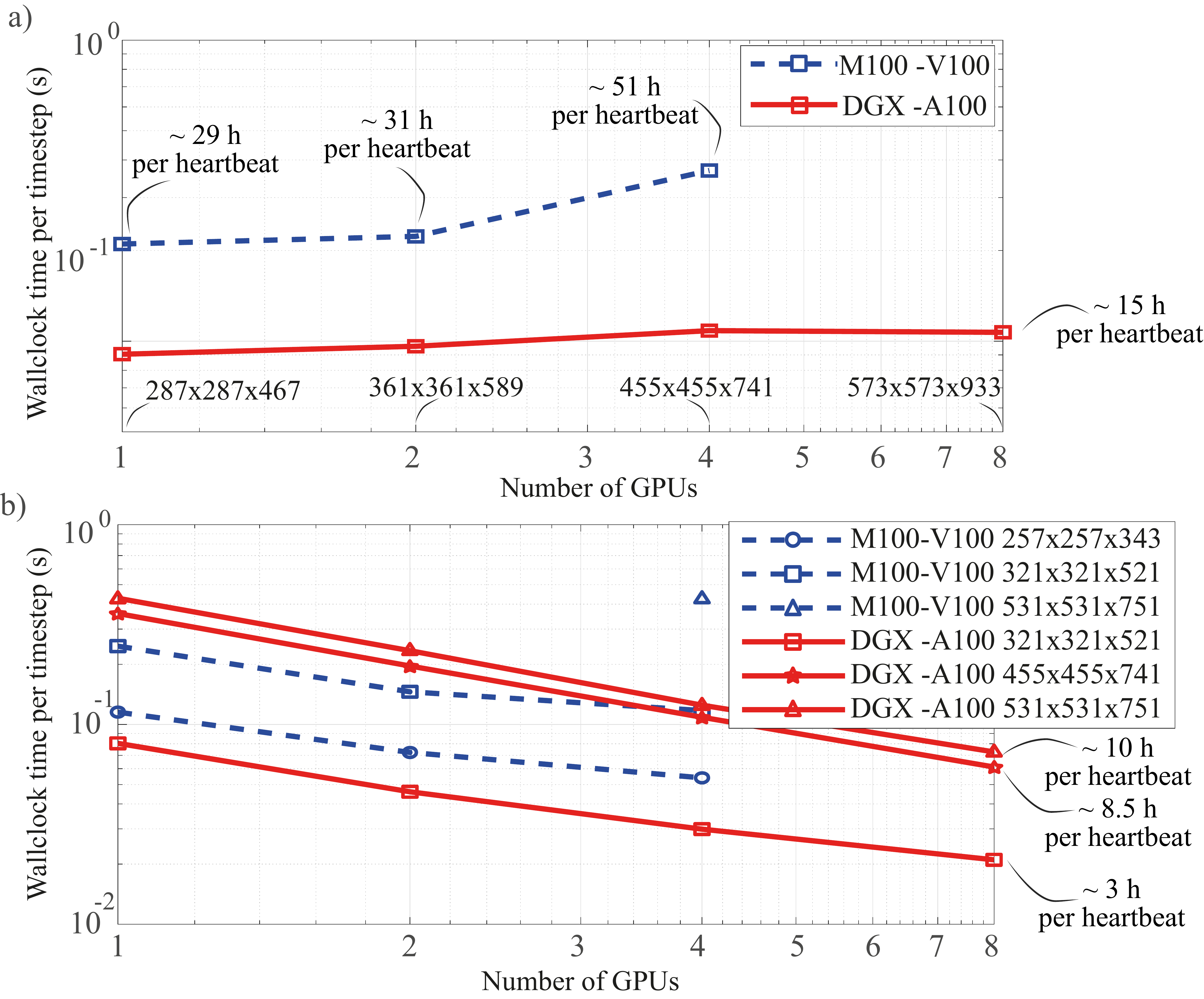}
\caption{(a) Weak and   b) strong scaling performance of the GPU accelerated code running on one node of Marconi100 (blue dashed curves) equipped with V100 cards and on a DGX machine (red solid curves) mounting A100 cards. The grid points $N_x \text{x} N_y \text{x} N_z$ are reported for the two periodic direction -- x and y -- and for the wall-normal one,  z. The time to integrate a whole heartbeat is obtained by scaling the wallclock time of a time step by the number of timesteps (taken equal to half of a  million) needed to integrate 1~second, which corresponds to a heart rate of 60~beats--per--minute.
 }
\label{fig:scaling}
\end{figure}
In order to test the computational efficiency of the GPU accelerated FSEI,  we have run the code on Marconi100 the GPU accelerated cluster from CINECA equipped with V100 cards and on the novel DGX machine from Nvidia mounting the next generation A100 cards. Rather than running on multiple nodes, the code performance has been tested on a single node since such a  limited computational hardware can be, in principle, hosted also in a hospital. 
For this analysis, we have initialized a left cardiac geometry as the one detailed in Section~\ref{sec:problem} using 40000 tetrahedrons for the 3D myocardium and 18000 triangles for  the 2D wet surfaces. However, as the Eulerian grid is refined, the Lagrangian resolution should be refined accordingly in order to ensure the correct enforcement of the no-slip boundary condition using the IB method,  and consequently, the number of tetrahedrons and triangles should be a function of the Eulerian grid at use. Such a constraint not only requires to remesh the whole cardiac geometry any time the grid of the fluid solver is refined, but it would also affect the scaling tests of the code as any Eulerian grid would correspond to a different Lagrangian one.
This issue can be avoided by using an adaptive Lagrangian mesh refinement procedure where the triangular mesh is automatically subdivided into smaller subtriangles (called tiles) until each one gets smaller than the local Eulerian grid size, thus avoiding 'holes' in the interfacial boundary condition. This way, the heart tissues can be discretized by adequately resolving the geometric details, but independently of the Eulerian mesh, and each triangle is successively refined until the Lagrangian resolution of the tiled grid is sufficiently high; we refer to \cite{fsei,fseichap} for a more comprehensive discussion of the method. As a result, the same tetrahedral and triangular meshes are used for all the scaling tests presented here and, as the Eulerian grid is refined, only the number of tiles where the no--slip condition is enforced increases, whereas the mesh used to solve the structural dynamics and the electrophysiology is unvaried. 
\\ \indent 
Figure~\ref{fig:scaling}(a) shows the wallclock time per time step as a function of the number of GPUs, with the number of grid points increasing proportionally to the number of computing cards starting from an initial grid of 287x287x467 corresponding to about 70\% of the available memory of a single V100 (16~GB). 
As the number of cards is increased from one to two, the computational time remains about the same with a wallclock time of about 0.2~s per time step, thus showing a good weak scaling  properties, although when the number of GPU cards and grid points are further doubled the computing time increases significantly. This worsening of the performance can be rationalized by recalling the architecture of the Marconi100 where each node is equipped with two pairs of GPUs and each pair mounted on the CPU sockets. 
Consequently, the cards within the same pair are connected by the fast NVLink~2.0 connection allowing for an efficient all--to--all communication among the slabs, whereas the two pairs of cards are connected by a slower 64 GBps X bus. The latter connection significantly reduces the speed of the all--to--all communications between the pairs of cards that is needed to solve both the equation for the provisional velocity and for the elliptic equation to impose mass conservation (see Section~\ref{sec:NS}), which become a bottleneck compromising the code scalability.
For this reason, the same weak scaling test has been  run on the novel Nvidia DGX machine equipped with 8 GPUs A100 all connected through the next generation NVLink~3.0 for a total GPU memory of  640 GB.
As indicated by the red line in Figure~\ref{fig:scaling}(a),  not only the computational time is reduced owing to the faster GPU cards, but also the weak scaling works satisfactory and the grid is increased from one  to eight cards preserving a wallclock time of about 0.1~s per time step.
\\ \indent 
In contrast to many turbulent flows where the Reynolds number in the simulation is limited by the computational resources available and by the weak scaling performance of the code, 
in cardiovascular flows it is pointless to increase the Reynolds number above the one fixed by the human physiology in healthy and pathologic conditions. In this framework, it is more relevant testing the code speedup for a given (converged) grid as more computational resources become available (strong scaling), rather than preserving the same computing time when both the grid refinement and the number of GPUs are increased (weak scaling as discussed above).
In Figure~\ref{fig:scaling}(b) the strong scaling results for different grids running on Marconi100 (blue lines) and on the DGX  machine (red lines) are shown. The smallest grid running on Marconi~100 (257x257x343 grid points in the three directions) corresponds to the one used for the production simulations used in the next section. Specifically, when the number of cards is doubled from one to two the wallclock time reduces from 0.115 to 0.072~s corresponding to a speedup of about 1.60, whereas doubling again from two to four cards it reduces to 1.34 owing to the slower communications among the GPU couples connected through the X bus, as explained above for the weak scaling. A similar behaviour is observed for the second grid considered, 312x321x521, corresponding to the allocation of about 90\% of the available memory of a single V100 GPU card on Marconi100. The same grid has been tested on the DGX machine yielding a significantly lower wallclock time, namely about $-67\%$ using 1 or 2 cards and $-74\%$ using 4 cards, owing to the new A100 GPU cards and the faster connection among cards, NVLink 3.0 among all the cards rather than NVLink 2.0  plus Xbus connection between the couples as on Marconi100.
Although the size of the grid limits the code speedup ranging from 1.7 (one card to two cards) to 1.4 (four cards to eight cards), it should be noted that the wallclock time using 8 A100 cards allows to solve a whole heartbeat for the left heart in about 3 hours, thus greatly reducing the time--to--solution needed to timely provide computational results to the medical doctor to aid clinical decisions. 
Nevertheless, a speedup exceeding 1.8 is observed for more refined grids such as 455x455x741 and 531x531x751, which correspond to 8.5 and 10 hours to integrate a single heartbeat. 
\\ \indent 
Remarkably, the more refined 531x531x751 grid corresponding to a memory allocation above  95 \% on four V100 cards (with a total available memory of 16Gb x4=64Gb) can be allocated on a single A100 card providing a wallclock time per time step similar to  the one measured using four V100 cards (0.42s and 0.43 s, respectively). The time to solution is then reduced by a factor -46\%, -71\% and -83\%, as the number of A100 cards is increased to 2, 4 and 8, respectively. 
\\ \indent 
It should be remarked that especially on these finer grids suitable for the whole heart modelling the GPU accelerated code results in significantly better computational performance with respect to the original CPU code accelerated through MPI and openMP.
Please see Appendix~A for the scaling tests of the CPU version of the code.

\section{Application: the left human heart}\label{sec:problem}
As a demonstration of the GPU accelerated FSEI, we show some results for the left heart of a healthy subject. The computational domain is similar to the one used in \cite{fseichap} which relied on the CPU version of the code: we refer to this reference for further details on the geometrical and electrophysiology parameters of the cardiac configuration.
The computational domain is sketched in Figure~\ref{fig:vangelder} and comprises a left atrium and ventricle that are discretized as a whole elastic three--dimensional (3D) medium with attached a set of slender bodies (hence modelled as 2D shells), namely the bileaflet mitral valve, the three-leaflet aortic valve and the thoracic aorta.  
The 2D structures are bound to the 3D structure or to another 2D structure so that to avoid, as an example, the aorta to  detach from the ventricle (or the aortic leaflets from the aorta) during the simulation. 
For each couple of fastened structures, a master and a slave is defined in the preprocessing 
and for each slave vertex to be bound (e.g. the tip of the thoracic aorta to the ventricle) the closest master vertex and the corresponding vector distance between them are determined and stored. During the simulation the instantaneous position of the binding slave vertices is thus forced to be equal to the vectorial sum of the instantaneous position of the  corresponding master vertex plus the initial vector distance found in the preprocessing. In particular, the 2D mitral valve leaflets and aorta are bound to the 3D myocardium, whereas the 2D aortic valve leaflets are bound to the aorta.
\\ \indent 
The reference frame is defined with the $z$-axis oriented as the longer ventricle axis and pointing down towards its apex, the $x-z$ plane is identified with the symmetry plane of the ventricle. The left ventricle has a stress-free volume of $125$~ml and is connected to the aorta through the aortic orifice with a diameter $d^a=19$~mm where the three-leaflet aortic valve is placed. On the other hand, the ventricle is connected to the atrium (with a free--stress volume of 40~ml) through a circular orifice of diameter $d^m=24$~mm where the bileaflet mitral valve is mounted. The Reynolds number  is defined using as reference length and velocity, the diameter of the mitral orifice and the average speed through the mitral annulus during diastole measured using Doppler echocardiography ($U^m=60$~cm/s): $Re=U^m d^m/\nu=3000$, with $\nu$ the effective kinematic viscosity for human blood with an hematocrit of $40\%$ (Newtonian blood model). The hemodynamics is thus solved in a Cartesian domain of size $l_x \times l_y \times l_z=96 \times 96 \times 156$~mm$^3$ with periodic conditions in the $x,y$ directions and no--slip Dirichlet condition on the velocity in the $z$ direction. 
The left heart is immersed in the fluid domain without intersecting the boundaries of the Eulerian grid and during its dynamics it can suck (propel) blood through the inlets (outlet) of the pulmonary veins (aorta) from (to) the outer blood volume, which serves as a numerical blood reservoir connected to the left heart at study.
Since the left heart is decoupled from the rest  of the circulatory system, a localized volume forcing at the pulmonary veins inlet and aorta outlet directed as the normal to the section towards the left heart is imposed so as to mimic the hydraulic resistance  of the vascular network not included in the computational domain, see \cite{fseichap}.
Even if any phase of the cycle could be used as initial condition, the beginning of the systole is the most convenient as the cardiac valves are closed, the heart chambers are in the stress-free configuration and only the aorta needs a pretensioning load.
\\ \indent  
The myocardium is modelled as a uniform conductive medium and the diection of the fast conductivity fibers has been accounted for by setting the conductive tensor $ \doubleunderline{M}^{int} $ and $ \doubleunderline{M}^{ext}$ so that the computational model reproduces the benchmark timings of ventricular and atrial depolarization. 
Since the sinoatrial node located in the upper part of the right atrium is not included in the computational domain, a localized triggering impulse, $I_s$, is prescribed with the appropriate delays at the Bachmann and His bundles, respectively, for the left atrium and ventricle. These electrical impulses trigger the muscle contraction and the time period between two consecutive input currents at the bundles is set equal to 1000~ms, which corresponds to a heart beat of 60~bpm and  a Womerseley number of $Wo = d^m / \sqrt{T^* \nu}=10.73$. 

\subsection{Electrical activation and muscle contraction}
\begin{figure}[h!]
\centering
\includegraphics[width=.9\textwidth]{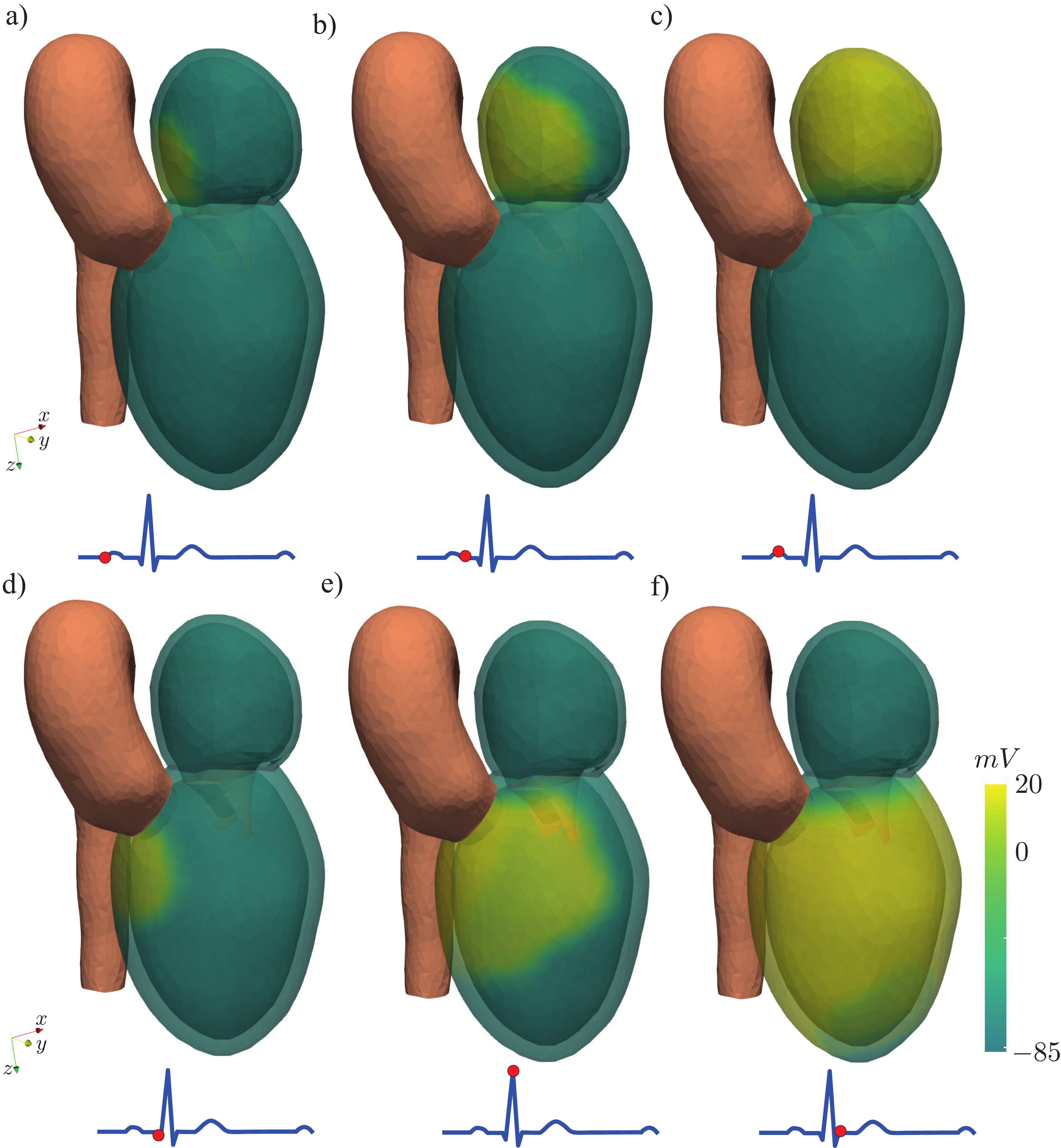}
\caption{Snapshots of the electrical activation of the left atrium, (left) front and (right) lateral view. The ECG profile indicates the phase within the heart beat. }
\label{fig:electroA}
\end{figure}
The electrical activation of the left atrium and ventricle is shown in Figure~\ref{fig:electroA} by visualizing the isocontours of the transmembrane potential, where the base--level (green isolevel) indicate the resting potential of about $90$~mV.
As visible in Figure~\ref{fig:electroA}(a), the electrical impulse applied at the Bachmann bundle induces a  local depolarization of the myocardium, which exhibits a positive transmembrane potential of about $20$~mV.  This local depolarization fosters the depolarization of the neighboring myocytes  and the resulting propagating wavefront travels across the atrial myocardium (panel \ref{fig:electroA}b), and the whole chamber is electrically activated after about 90~ms (panel \ref{fig:electroA}c), which corresponds to the P wave in the  electrocardiogram (ECG).
A similar behaviour is observed for the ventricular activation corresponding to the so--called QRS--complex in the ECG and lasting about 100~ms. The electrical impulse originated at the His bundle (panel \ref{fig:electroA}d) locally depolarizes the ventricular myocardium (\ref{fig:electroA}e)  then spreads around the atrioventricular node until (\ref{fig:electroA}f) the whole ventricle is activated.

\subsection{Cardiac hemodynamics}\label{sec:hemo}
 \begin{figure}[h!]
\centering
\includegraphics[width=1.1\textwidth]{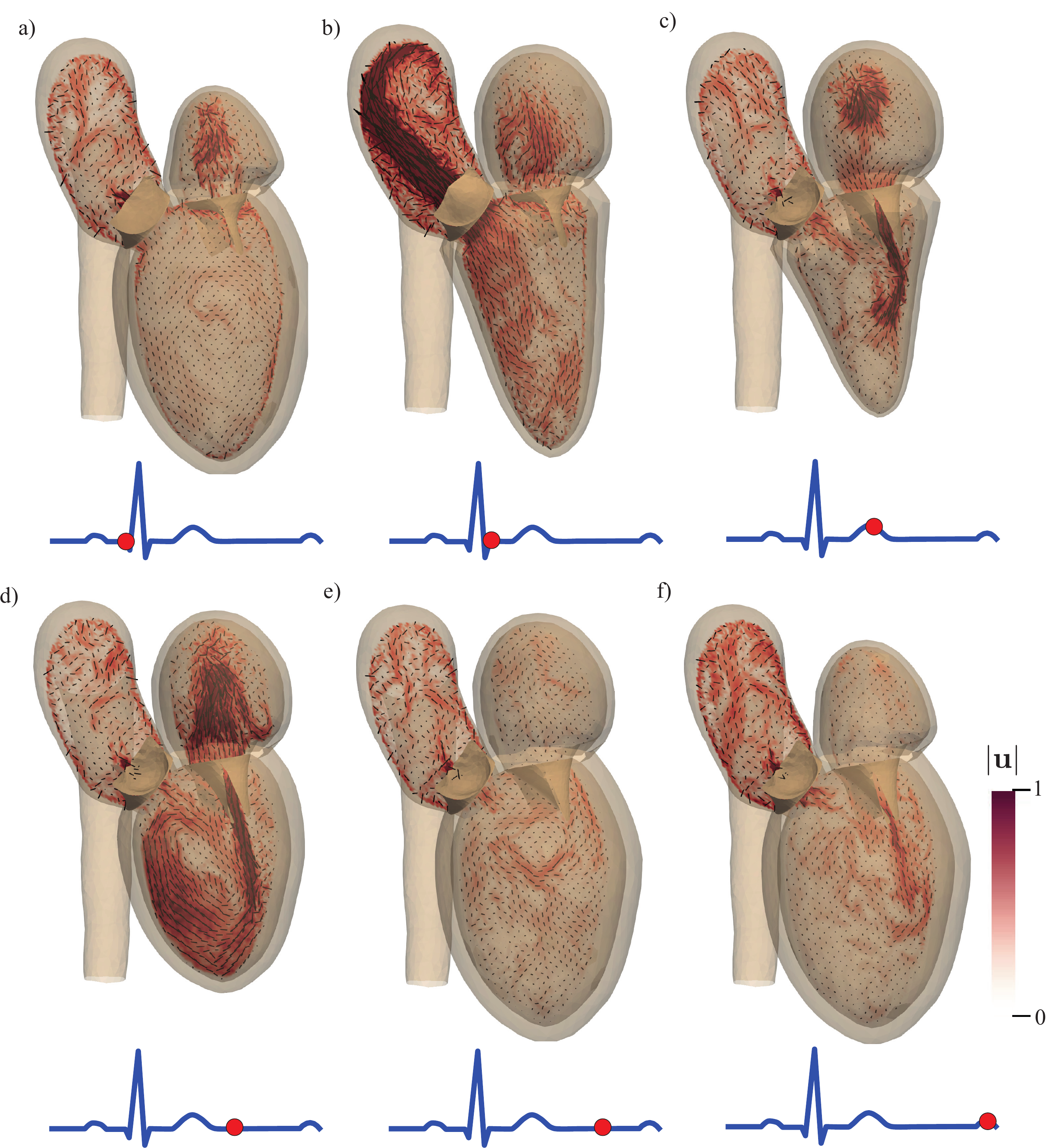}
\caption{Instantaneous snapshots of the nondimensional velocity vectors in the $x-z$ symmetry plane and contours of the velocity magnitude colored by the velocity magnitude times the sign of the vertical velocity. The ECG profile indicates the phase within the heart beat. }
\label{fig:pack_vel}
\end{figure}
Figure~\ref{fig:pack_vel} shows the corresponding hematic flow in the symmetry plane ($x-z$) driven by the muscular contraction, where the heart phase is indicated within a typical ECG profile and the velocity vectors are superimposed on the isocontours of the velocity magnitude.
At beginning systole (panel \ref{fig:pack_vel}a), the ventricular pressure increases, and an incipient regurgitation is observed through the mitral channel before the blood is ejected into the aorta through the aortic channel as depicted in (\ref{fig:pack_vel}b).  During early diastole (\ref{fig:pack_vel}c), the ventricle relaxes, and the hematic flow accelerates through the mitral orifice thus opening the valve, and  as a consequence, a strong mitral jet is produced, which  is initially directed towards the ventricle lateral wall owing to the asymmetry of the leaflets. This initial rapid filling of the ventricle owing to the elastic restoring force is called the E--wave, and when peak blood flux into the ventricle is attained (\ref{fig:pack_vel}d), the leaflets open wider, the jet points vertically down to the ventricle apex and a single large vorticity structure takes place occupying the whole ventricle in agreement with diastolic measurements both in--vivo and in--vitro \cite{Fortini2013,viola2019left}. 
After the initial passive filling has slowed down then (\ref{fig:pack_vel}e) diastasis starts and the main vorticity structure decays before (\ref{fig:pack_vel}f) another fluid injection called the A--wave is generated by the atrial systole creating a second mitral jet, but weaker. At the end of the diastole, the initial configuration is recovered and the cardiac cycle repeats itself.

\subsection{Wiggers diagram}
\begin{figure}[t!]
\centering
\includegraphics[width=1\textwidth]{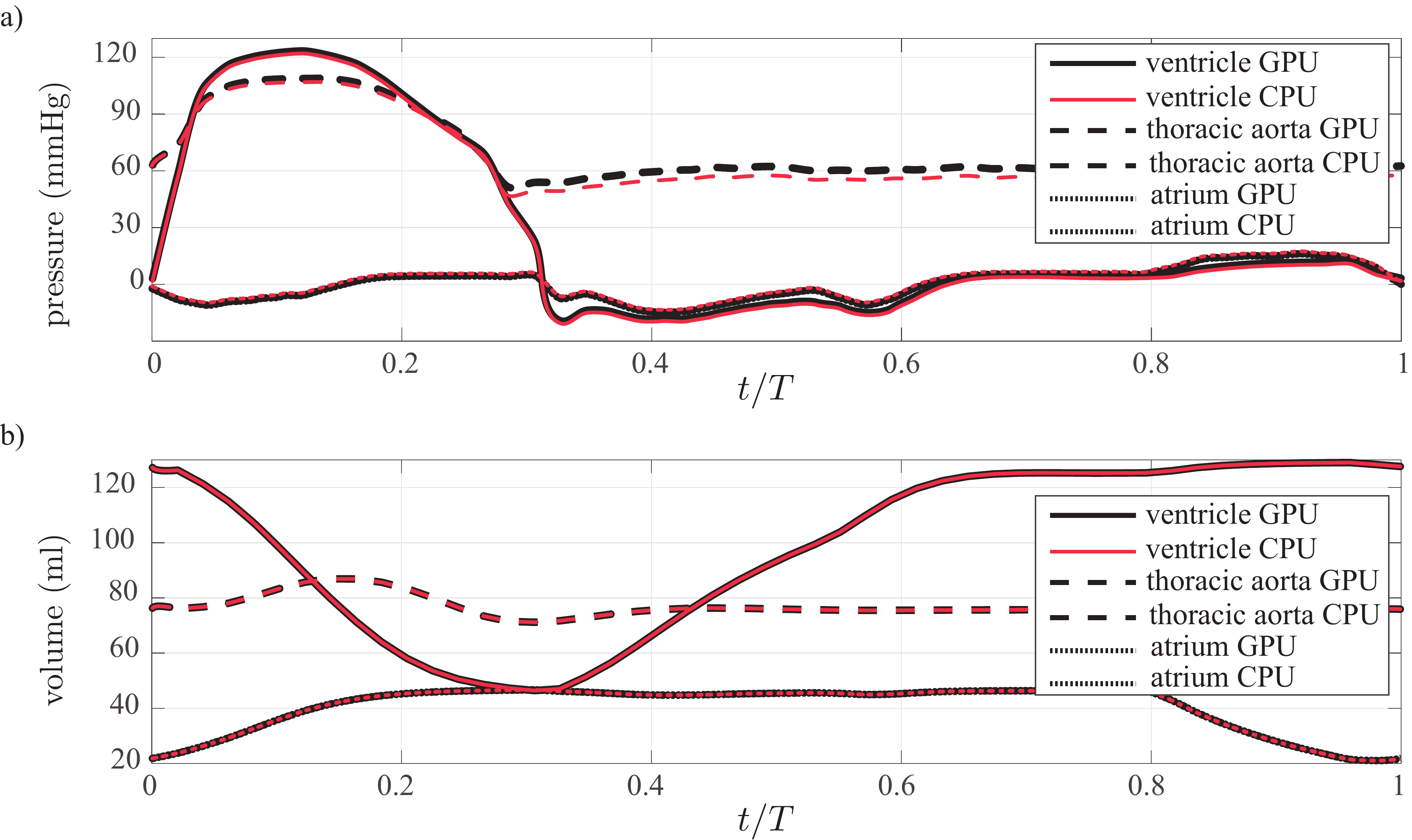}
\caption{Wiggers diagram \cite{hall2010guyton} obtained by the numerical simulation showing (a) the time evolution of the ventricular, atrial and aortic (thoracic tract) pressures, along with (b) the ventricular, atrial, aortic (thoracic tract) volumes as a function of time normalized by the beating period, $t/T$.  Black (red) lines correspond to GPU (CPU) simulations, all quantities have been phase--averaged over five heart beats.}
\label{fig:wiggers}
\end{figure}

The variations in pressure and volume described above can be portrayed in the Wiggers diagram that is a standard representation of the heart physiology. In order to show the agreement between the CPU-- and the GPU-- compiled codes, both curves are shown in Figure~\eqref{fig:wiggers}, where the pressure and volume of the heart chambers have been phase--averaged over five heart beats. 
The ventricular contraction triggered by the electrical discharge of the myocardium  (QRS in the ECG) causes the ventricular pressure to increase, which exceeds the one in the thoracic aorta, thus opening the semilunar valve squeezing the blood from the ventricle into the aorta. After the ventricular volume gets to its minimum (end--systolic--volume ESV), the ventricular muscle starts relaxing producing a fall of the ventricular pressure and the semilunar valve closes. Simultaneously, diastole starts, and the ventricular filling begins when the mitral valve opens and the volume of the former rapidly increases. Ventricular pressure remains low during this filling, whereas the ventricular volume reaches the stress--free volume and further increases a little more when the atria contract (end--diastolic--volume EDV). This contraction causes a small increase in atrial and ventricular pressures (and is associated with the P wave of the ECG).
The corresponding \textit{stroke volume} normalized by the EDV expressed as a percentage provides the \textit{ejection fraction},
$
EF = \frac{ SV }{EDV}\%= \frac{81.0~\text{ml}}{127.6~\text{ml}} = 63.5\%,
$
which is  a measure of the efficiency of the heart functioning, with healthy values for a normal subject in between $50\%$ and $70\%$. In the case of a heart beating at 60~bpm, as investigated here, the cardiac output is equal to 
$
CO = SV \times HR = 81.0~\text{ml} \times 60~\text{bpm} =4.86~\text{l/min},
$
which is a typical physiological value for the heart of a healthy adult.

\section{Conclusions}
The FSEI is a promising tool to provide a prediction on the patient hemodynamics. However, running the whole FSEI is computationally expensive as three solvers have to be used simultaneously and the relatively high Reynolds number (about 3000 based on the diameter of the mitral orifice and the peak intraventricular velocity) together with the stiffness of the myocardium introduce short time scales, thus calling for fine grids and small time steps. Indeed, about half of a million of time steps are needed to solve a single heartbeat and, consequently, at least 5 millions time steps have to be advanced to integrate 10 heartbeats and obtain phase--averaged data. Clearly, these calculations can not be executed on a small desktop computer and should be tackled using high--performance computing facilities to reduce the time to solution. On the other hand, such an approach would be of little use in the clinical practice as medical doctors would  need to wait days for the results coming from a remote computing facility before deciding about the patient prognosis. 
\\ \indent In this work, a GPU acceleration of the FSEI code has been developed with the aim of making the code as efficient as possible to run using on premises computing resources composed of a few GPUs cards while maintaining time to solution within hours. Indeed, GPUs are a compact numerical engine optimized to execute a large number of threads in parallel, which is a crucial point for a systematic use of cardiovascular flow simulations in the clinical practice.
To this aim, the initial CPU code parallelized using MPI, has been ported in CUDA Fortran with the extensive use of kernel loop directives (CUF kernels) in order to have a source code as close as possible to the original CPU version.
The resulting  GPU accelerated multi-physics heart model shows good strong scaling characteristics, and the wall-clock time per step for the GPU version is in between one and two orders of magnitude smaller than that of  the CPU code thus allowing for a timely solution of the intraventricular hemodynamics. 
\\ \indent
Our computational environment can simulate several patient heartbeats overnight, thus timely providing the phase--averaged results to the medical doctor and, importantly, the required hardware can fit in an office or in a dedicated computing facility in a clinic or in a hospital. The FSEI scaling performance has been tested both using the V100 and the faster A100 GPU cards and the speedup documented here on the DGX Nvidia machine will be  obtained with the upcoming pre-exascale supercomputers and is expected to further improve with the next generation cards. The GPU-accelerated FSEI introduced in this paper is thus a further step towards the development of physical and CFD aided medical diagnostic to investigate pathologies and test surgical procedures.

\section*{Appendix A: scaling of the CPU version of the code}
\begin{figure}[t!]
\centering
\includegraphics[width=.75\textwidth]{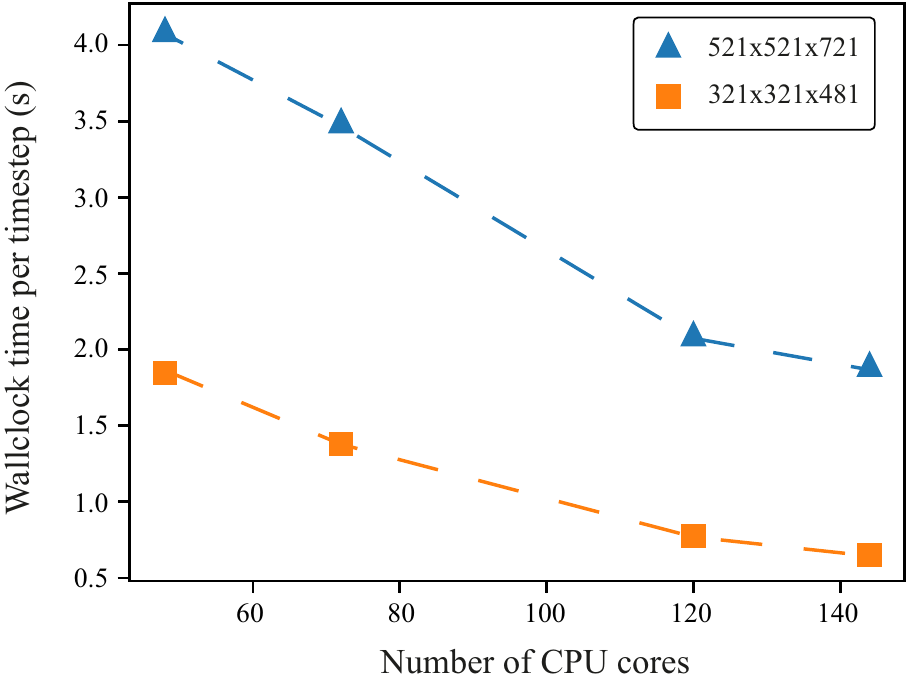}
\caption{Strong scaling performance of the CPU code for a grid of $Nx=513,~Ny=513~\text{and}~Nz=721$ nodes and another of $Nx=321,~Ny=321~\text{and}~Nz=481$, with the fluid solver parallelized using MPI directives.}
\label{fig:scalingCPU}
\end{figure}

As a final analysis, the CPU strong scaling is reported in Figure~\ref{fig:scalingCPU} using a grid of 531x531x721 nodes running on Cartesius from the Dutch Supercomputing Consortium SURFsara equipped with 2x12--core 2.6 GHz Intel Xeon E5-2690 v3 Haswell nodes with 64 GB of memory per node and 56 Gbit/s inter-node FDR InfiniBand. The CPU code provides the same results as the GPU version up to machine precision (see 
Figure~\ref{fig:wiggers}) with a reasonable strong scaling. However, the wall clock times per iteration across different resolutions is high compared to the results achieved on GPU's which eliminates the capability to simulate multiple heart beats using pure CPU parallelization in an economic manner.  

\section*{Acknowledgments}
This work has been partly supported with the
865 Grant 2017A889FP 'Fluid dynamics of hearts at risk of failure: towards methods for the prediction of disease progressions' funded by the Italian Ministry of Education and University.

\section*{References}

\bibliography{mybibfile}

\begin{thebibliography}{10}
\expandafter\ifx\csname url\endcsname\relax
  \def\url#1{\texttt{#1}}\fi
\expandafter\ifx\csname urlprefix\endcsname\relax\def\urlprefix{URL }\fi
\expandafter\ifx\csname href\endcsname\relax
  \def\href#1#2{#2} \def\path#1{#1}\fi

\bibitem{fsei}
F.~Viola, V.~Meschini, R.~Verzicco, Fluid--structure-electrophysiology
  interaction (fsei) in the left-heart: A multi-way coupled computational
  model, European Journal of Mechanics-B/Fluids 79 (2020) 212--232.

\bibitem{sackett1997evidence}
D.~L. Sackett, Evidence-based medicine, in: Seminars in perinatology, Vol.~21,
  Elsevier, 1997, pp. 3--5.

\bibitem{Fedosov}
D.~A. Fedosov, B.~Caswell, G.~E. Karniadakis, Systematic coarse-graining of
  spectrin-level red blood cell models, Computer Methods in Applied Mechanics
  and Engineering 199~(29-32) (2010) 1937--1948.

\bibitem{hammer2011mass}
P.~E. Hammer, M.~S. Sacks, J.~Pedro, R.~D. Howe, Mass-spring model for
  simulation of heart valve tissue mechanical behavior, Annals of biomedical
  engineering 39~(6) (2011) 1668--1679.

\bibitem{tung1978}
L.~Tung, A bi-domain model for describing ischemic myocardial dc potentials.,
  Ph.D. thesis, Massachusetts Institute of Technology (1978).

\bibitem{ten2006}
K.~ten Tusscher, A.~Panfilov, Cell model for efficient simulation of wave
  propagation in human ventricular tissue under normal and pathological
  conditions, Physics in Medicine \& Biology 51~(23) (2006) 6141.

\bibitem{cudabook}
G.~Ruetsch, M.~Fatica, CUDA Fortran for scientists and engineers: best
  practices for efficient CUDA Fortran programming, Elsevier, 2013.

\bibitem{Meschini2018}
V.~Meschini, M.~De~Tullio, G.~Querzoli, R.~Verzicco, Flow structure in healthy
  and pathological left ventricles with natural and prosthetic mitral valves,
  Journal of fluid mechanics 834 (2018) 271--307.

\bibitem{fseichap}
F.~Viola, V.~Meschini, R.~Verzicco, A computational tool for unprecedented
  simulations of the left cardiac functioning: an electro--fluid--structure
  solver, under revision in the book \textit{Topics in biomechanics of
  cardiovascular diseases}, Springer Series in Solid and Structural Mechanics.

\bibitem{Katritsis2007}
D.~Katritsis, L.~Kaiktsis, A.~Chaniotis, J.~Pantos, E.~P. Efstathopoulos,
  V.~Marmarelis, Wall shear stress: theoretical considerations and methods of
  measurement, Progress in cardiovascular diseases 49~(5) (2007) 307--329.

\bibitem{DeVita}
F.~De~Vita, M.~D. de~Tullio, R.~Verzicco, Numerical simulation of the
  non-newtonian blood flow through a mechanical aortic valve, Theoretical and
  Computational Fluid Dynamics 30(1) (2016) 129--138.

\bibitem{raimoin}
M.~Rai, P.~Moin, Direct simulations of turbulent flow using finite-difference
  schemes, in: 27th Aerospace Sciences Meeting, 1991, p. 369.

\bibitem{Verzicco1996}
R.~Verzicco, P.~Orlandi, A finite-difference scheme for three-dimensional
  incompressible flows in cylindrical coordinates, Journal of Computational
  Physics 123~(2) (1996) 402--414.

\bibitem{pencil}
E.~P. van~der Poel, R.~Ostilla-M{\'o}nico, J.~Donners, R.~Verzicco, A pencil
  distributed finite difference code for strongly turbulent wall-bounded flows,
  Computers \& Fluids 116 (2015) 10--16.

\bibitem{uhlmann2005immersed}
M.~Uhlmann, An immersed boundary method with direct forcing for the simulation
  of particulate flows, Journal of Computational Physics 209~(2) (2005)
  448--476.

\bibitem{Vanella2009}
M.~Vanella, E.~Balaras, A moving--least--squares reconstruction for
  embedded--boundary formulations, Journal of Computational Physics 228(18)
  (2009) 6617--6628.

\bibitem{TuPa}
M.~D. De~Tullio, Pascazio, A moving-least-squares immersed boundary method for
  simulating the fluid-structure interaction of elastic bodies with arbitrary
  thickness, to appear in Journal of Computational Physics.

\bibitem{kim1985application}
J.~Kim, P.~Moin, Application of a fractional-step method to incompressible
  navier-stokes equations, Journal of computational physics 59~(2) (1985)
  308--323.

\bibitem{vangelder}
A.~V. Gelder, Approximate simulation of elastic membranes by triangulated
  spring meshes, Journal of graphics tools 3~(2) (1998) 21--41.

\bibitem{costa1996three}
K.~D. Costa, P.~J. Hunter, J.~Wayne, L.~Waldman, J.~Guccione, A.~D. McCulloch,
  A three-dimensional finite element method for large elastic deformations of
  ventricular myocardium: Ii--prolate spheroidal coordinates.

\bibitem{usyk2000effect}
T.~Usyk, R.~Mazhari, A.~McCulloch, Effect of laminar orthotropic myofiber
  architecture on regional stress and strain in the canine left ventricle,
  Journal of elasticity and the physical science of solids 61~(1-3) (2000)
  143--164.

\bibitem{kantor1987}
Y.~Kantor, D.~R. Nelson, Phase transitions in flexible polymeric surfaces,
  Physical Review A 36~(8) (1987) 4020.

\bibitem{li2005}
J.~Li, M.~Dao, C.~Lim, S.~Suresh, Spectrin-level modeling of the cytoskeleton
  and optical tweezers stretching of the erythrocyte, Biophysical journal
  88~(5) (2005) 3707--3719.

\bibitem{Clayton2008}
R.~H. Clayton, A.~V. Panfilov, A guide to modelling cardiac electrical activity
  in anatomically detailed ventricles, Progress in biophysics and molecular
  biology 96~(1) (2008) 19--43.

\bibitem{sundnes2007}
J.~Sundnes, G.~T. Lines, X.~Cai, B.~F. Nielsen, K.-A. Mardal, A.~Tveito,
  Computing the electrical activity in the heart, Vol.~1, Springer Science \&
  Business Media, 2007.

\bibitem{moukalled2016finite}
F.~Moukalled, L.~Mangani, M.~Darwish, et~al., The finite volume method in
  computational fluid dynamics, Vol. 113, Springer, 2016.

\bibitem{trefethen1997numerical}
L.~N. Trefethen, D.~Bau~III, Numerical linear algebra, Vol.~50, Siam, 1997.

\bibitem{rush1978practical}
S.~Rush, H.~Larsen, A practical algorithm for solving dynamic membrane
  equations, IEEE Transactions on Biomedical Engineering~(4) (1978) 389--392.

\bibitem{marsh2012secrets}
M.~E. Marsh, S.~T. Ziaratgahi, R.~J. Spiteri, The secrets to the success of the
  rush--larsen method and its generalizations, IEEE transactions on biomedical
  engineering 59~(9) (2012) 2506--2515.

\bibitem{nash2004}
M.~P. Nash, A.~V. Panfilov, Electromechanical model of excitable tissue to
  study reentrant cardiac arrhythmias, Progress in biophysics and molecular
  biology 85~(2-3) (2004) 501--522.

\bibitem{afidgpu}
X.~Zhu, E.~Phillips, V.~Spandan, J.~Donners, G.~Ruetsch, J.~Romero,
  R.~Ostilla-M{\'o}nico, Y.~Yang, D.~Lohse, R.~Verzicco, et~al., Afid-gpu: a
  versatile navier--stokes solver for wall-bounded turbulent flows on gpu
  clusters, Computer physics communications 229 (2018) 199--210.

\bibitem{vamsi2017}
V.~Spandan, V.~Meschini, R.~Ostilla-M{\'o}nico, D.~Lohse, G.~Querzoli, M.~D.
  de~Tullio, R.~Verzicco, A parallel interaction potential approach coupled
  with the immersed boundary method for fully resolved simulations of
  deformable interfaces and membranes, Journal of computational physics 348
  (2017) 567--590.

\bibitem{vamsi2018}
V.~Spandan, D.~Lohse, M.~D. de~Tullio, R.~Verzicco, A fast moving least squares
  approximation with adaptive lagrangian mesh refinement for large scale
  immersed boundary simulations, Journal of computational physics 375 (2018)
  228--239.

\bibitem{Fortini2013}
S.~Fortini, G.~Querzoli, S.~Espa, A.~Cenedese, Three-dimensional structure of
  the flow inside the left ventricle of the human heart, Experiments in fluids
  54~(11) (2013) 1--9.

\bibitem{viola2019left}
F.~Viola, E.~Jermyn, J.~Warnock, G.~Querzoli, R.~Verzicco, Left ventricular
  hemodynamics with an implanted assist device: An in vitro fluid dynamics
  study, Annals of biomedical engineering (2019) 1--16.

\bibitem{hall2010guyton}
J.~E. Hall, Guyton and Hall textbook of medical physiology, Elsevier Health
  Sciences, 2010.

\end{thebibliography}

\end{document}